\documentclass{iaus}
\usepackage{graphicx,natbib,aas_macros}

\title[Supernova-driven dynamo]%
{Supernova-driven interstellar turbulence\\ and the galactic dynamo}

\author[O.~Gressel, D.~Elstner \& G.~R{\"u}diger]%
{Oliver Gressel$^1$, Detlef Elstner$^2$ \& G{\"u}nther R{\"u}diger$^2$}

\affiliation{ $^1$Astronomy Unit, Queen Mary, University of London, %
              Mile End Road, London E1 4NS, UK \break
              $^2$Astrophysikalisches Institut Potsdam,  %
              An der Sternwarte 16, 14482 Potsdam, Germany %
  \break email:~\texttt{o.gressel@qmul.ac.uk, elstner@aip.de, gruediger@aip.de}}

\pubyear{2010}
\volume{274}
\pagerange{000--000}
\date{Sep. 9, 2010 and in revised form ??}
\jname{Advances in Plasma Astrophysics}
\editors{A. Bonanno, E. de Gouveia Dal Pino \& A. Kosovichev, eds.}

\newcommand\cm{\,\rm cm}

\newcommand\K{\,\rm K}
\newcommand\Myr{\,\rm Myr}

\newcommand\kms{\,\rm km\,s^{-1}}

\newcommand\pc{\,\rm\,pc}
\newcommand\kpc{\,\rm kpc}

\newcommand\tms{\!\times\!}

\newcommand{\mn}[1]{\overline{#1}}

\newcommand\mU{\mn{\mathbf{u}}}

\newcommand\mB{\mn{\mathbf{B}}}
\newcommand\EMF{\mathcal{E}}
\newcommand\Pm{\mathrm{Pm}}

\begin{document}

\maketitle

\begin{abstract}
The fractal shape and multi-component nature of the interstellar
medium together with its vast range of dynamical scales provides one
of the great challenges in theoretical and numerical astrophysics.
Here we will review recent progress in the direct modelling of
interstellar hydromagnetic turbulence, focusing on the role of energy
injection by supernova explosions. The implications for dynamo theory
will be discussed in the context of the mean-field approach.

Results obtained with the test field-method are confronted with
analytical predictions and estimates from quasilinear theory. The
simulation results enforce the classical understanding of a turbulent
Galactic dynamo and, more importantly, yield new quantitative
insights. The derived scaling relations enable confident global
mean-field modelling.

\end{abstract}

\firstsection 


\section{Interstellar turbulence}

Apart from stars, the baryonic matter within the Galaxy is in the form
of an extremely dilute, turbulent plasma known as the interstellar
medium (ISM). The multitude of physical processes within the ISM
entails a rich heterogeneous structure \citep{1978ppim.book.....S}.

Approximating radiative processes by a simplified cooling
prescription, and restricting the computational domain to a local
patch, the turbulent ISM is now routinely modelled by means of
three-dimensional fluid simulations
\citep[e.g.,][]{1999ApJ...514L..99K,2004A&A...424..817M,2005MNRAS.356..737S,%
2006ApJ...653.1266J,2006ApJ...638..797D}. One main focus of these
simulations has been to obtain filling factors of the different ISM
phases and compare them to the classical predictions as well as
observations \citep[e.g.,][]{1992FCPh...15..143D}. Further topics of
interest include turbulent mixing \citep{2005ApJ...634..390B},
thermodynamic distribution functions \citep{2005ApJ...626..864M}, and
line-of-sight integrated column densities \citep{2005ApJ...634L..65D}.

\subsection{The small-scale dynamo}

While various simulations
\citep{1999A&A...350..230K,2005A&A...436..585D,2005ApJ...626..864M}
discuss the influence of magnetic fields on the ISM morphology, little
is said about the actual mechanism of field
amplification. \citet{2004ApJ...617..339B} have addressed this
question by means of unstratified simulations of SNe turbulence. The
authors relate the growth of small-scale magnetic fields to vorticity
production in supernova shocks \citep{2001ApJ...563..800B}, and
chaotic field line-stretching \citep{2005ApJ...634..390B}.

The fact that vorticity production by colliding shells is almost
inevitable in a clumpy and highly structured ISM has first been
pointed out by \citet{1999A&A...350..230K}. The issue has then been
investigated for the simplified case of driven expansion waves by
\citet{2006MNRAS.370..415M} and, more recently, by Del Sordo \&
Brandenburg (this volume). Considering turbulence driven by
non-helical transverse waves, \citet{2004MNRAS.353..947H} have shown
that the small-scale dynamo becomes harder to excite in the
super-sonic regime, albeit the critical Reynolds number for the onset
of dynamo action only seems to depend weakly on the Mach number.

Because the eddy turnover time is short at small scales, a dynamo
based on chaotic field line stretching will be fast. This is in-line
with observations \citep[see][and this volume]{1996ARA&A..34..155B},
which exhibit dominant turbulent fields. Open issues remain with
respect to the mechanism governing the saturation of the small-scale
dynamo. Therefore, it is currently unclear whether equipartition field
strengths can be obtained by a non-helical dynamo alone.
Alternatively, the turbulent field might be explained as a
``shredded'' coherent field, i.e., as the by-product of a helical
mean-field dynamo.


\section{The large-scale galactic dynamo}

Notwithstanding the above, the presence of the observed coherent
fields on scales larger than the outer scale of the interstellar
turbulence \citep[$\sim 50\pc$, see][]{2010arXiv1001.5230F} clearly
requires the presence of a coherent dynamo. The favoured candidates
for such a dynamo are, in no particular order: (i) the kinetic driving
by SNe
\citep{1999ApJ...514L..99K,2008A&A...486L..35G,2009MNRAS.394L..84G},
(ii) the buoyant cosmic ray-supported Parker instability
\citep{1992ApJ...401..137P,2004ApJ...605L..33H}, (iii) the
magneto-rotational instability
\citep[MRI,][]{1999ApJ...511..660S,2004A&A...423L..29D,%
2006ApJ...641..862N,2007ApJ...663..183P}, and (iv) gravitational
interactions \citep{2010ApJ...716.1438K}.\footnote{Also see the
respective reviews by Hanasz, Otmianowska-Mazur, and Lesch (this
volume).}

The last of these effects is certainly dominant at the early stages of
galaxy formation and will provide a seed field for subsequent
processes. It remains to be shown, however, how important external
interactions are in the presence of a realistic feedback from scales
currently unresolved in cosmological simulations. Moreover, as
\citet{2009A&A...498..335H} have shown, the cosmic ray (CR) dynamo
critically relies on the anisotropy of the CR diffusion
coefficient. Combined simulations, including both SNe and CRs, will
prove whether this anisotropy remains effective for strongly tangled
turbulent fields.

The MRI will be important (at least) in the outer regions of galaxies,
where the star formation activity is low \citep{2010AN....331...34K}
-- even under moderate turbulence, it may operate efficiently. The
stability criterion for a global isothermal disk of thickness $2H$
leads to the relation
\begin{equation}
  8 \sqrt{\Pm_{\rm t}} < C_\Omega \equiv \frac{\Omega H^2}{\eta_{\rm t}}
\end{equation}
for instability \citep{2004A&A...424..565K}. Depending on the local
rotation frequency, $\Omega$, turbulent diffusivity $\eta_{\rm t}$,
and turbulent magnetic Prandtl number $\Pm_{\rm t}$ -- which is
generally expected to be of order unity \citep{2009A&A...507...19F} --
this value may be lower than the critical dynamo number for the
supernova-driven dynamo. It should be worthwhile to address this
question within state-of-the-art simulations of the ISM, with an
Alvf{\'e}n velocity of the external vertical field of the order
$1\kms$.

\begin{figure}
  \center\includegraphics[width=\columnwidth]{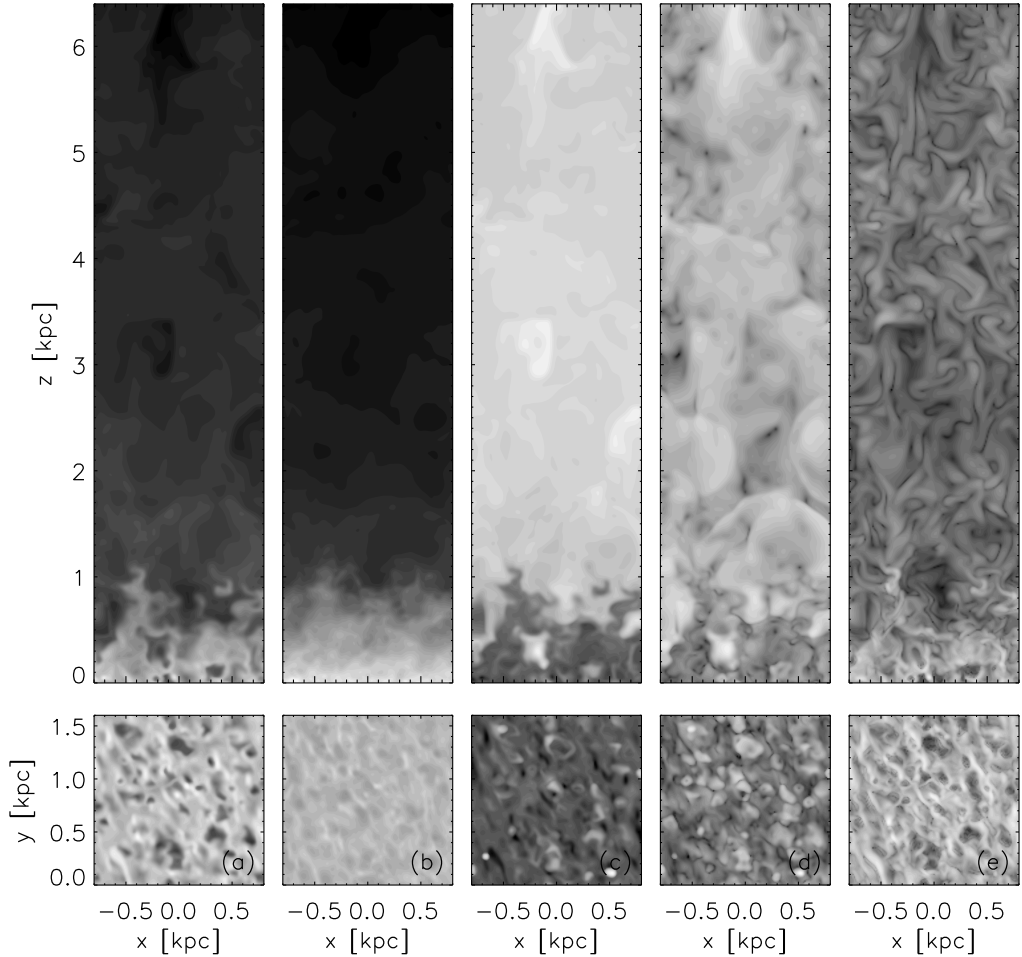}
  \caption{Snapshots at time $t=72\Myr$ of the top half of the now
        larger simulation box (\emph{upper panels}), and the disc
        midplane (\emph{lower panels}). The variables shown are: (a)
        number density $[\cm^{-3}]$, (b) column density $[\cm^{-2}]$,
        (c) temperature $[\K]$, (d) velocity dispersion $[\kms ]$, and
        (e) magnetic field strength $[\mu G]$.  The logarithmic grey
        scales cover ranges $[-4.76, 1.01]$, $[17.56,21.83]$, $[ 2.13,
        7.03]$, $[-0.61, 2.64]$, and $[-5.98,-1.18]$, respectively.}
  \label{fig:slices}
\end{figure}

\subsection{The supernova-driven dynamo}

The energy input through SNe into the ISM is tremendous. The
corresponding amplitude of the expected dynamo effect has first been
estimated by \citet{1990Natur.347...51S}. Assuming hydrostatic
equilibrium and applying quasi-linear theory
\citep{1993A&A...269..581R}, these estimates have subsequently been
refined by \citet{1996A&A...311..451F}. A shortcoming of the approach
was the neglect of a possible galactic wind. In general, there was a
controversy as to whether the turbulence created by the SNe was too
vigorous to warrant dynamo action -- either because of a too strong
wind (carrying the field away), or contrary to this, because of a too
strong downward pumping (enhancing turbulent dissipation near the
midplane).

In a series of papers, \citeauthor{1998A&A...335..488F} analytically
derived the electromotive force stemming from isolated remnants; the
line of work was also later supported by simulations of single
remnants \citep{1993A&A...274..757K,1996A&A...305..114Z}. To obtain
the net $\alpha$~effect, a convolution with an assumed vertical SN
distribution was applied. However, the approach suffered from a too
weak dynamo and highly dominant (upward) pumping. Only when
considering the stratified nature of the galactic disc
\citep{1998A&A...335..488F}, the issue was somewhat alleviated,
although still predicting a strong upward pumping.

Pioneering semi-global simulations based on ``first principles'' where
performed by \citet*{1999ApJ...514L..99K}, and it was only for the
fact of a too low dynamo number that no field amplification was
observed in their simulations. The first direct simulations exhibiting
dynamo action were reported almost a decade later
\citep{2008A&A...486L..35G}. The general morphology of such
simulations is illustrated in Figure~\ref{fig:slices}; note the
apparent correlation between the density~(a) and magnetic field
amplitude~(e) near the midplane, indicating field amplification via
compression. In contrast, away from the midplane (i.e., in the diffuse
medium) the field shows more folded structures.

\begin{figure}
  \center\includegraphics[width=\columnwidth]{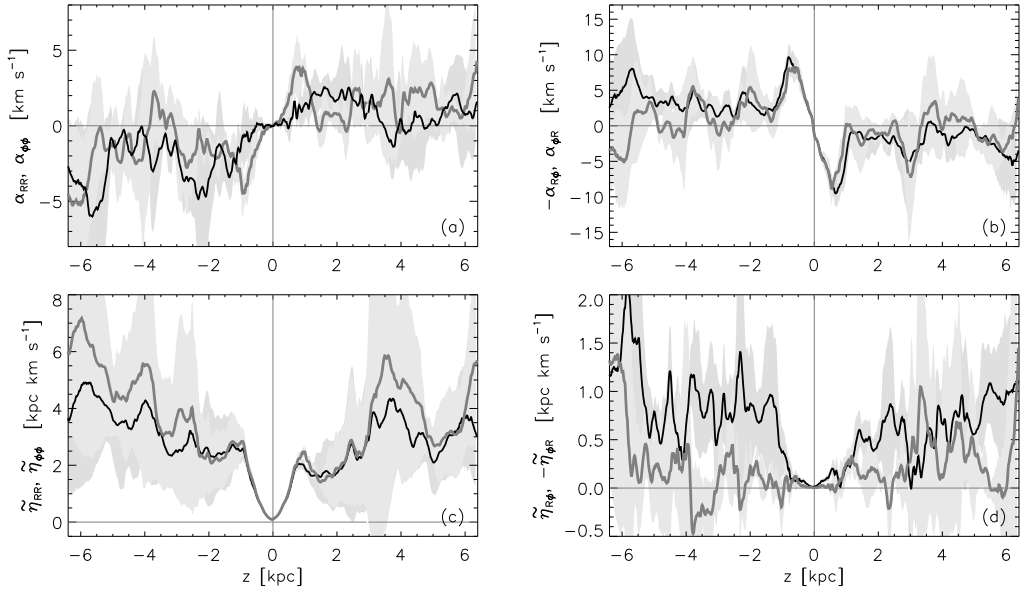}
  \caption{Dynamo coefficients as a function of the height $z$. The
    Variables are plotted in dark ($\alpha_{RR},\dots$) or light
    ($\alpha_{\phi\phi},\dots$) colours, respectively. Shaded areas
    indicate $1\sigma$-fluctuations. In the region for $|z|<2\kpc$,
    the results agree well with a previous run applying a much smaller
    box.}
  \label{fig:ae}
\end{figure}

The fast growth of the dynamo can be understood rigorously in terms of
mean-field theory, i.e. via a Reynolds-averaged induction equation
\begin{equation}
  \partial_{t}\mB = \nabla\tms\mn{\EMF} + \nabla \times 
                    \left[\ \mU\tms\mB  - \eta \nabla\tms\mB\ \right]\,,
  \label{eq:mf_ind}
\end{equation}
and assuming a standard parametrisation
\begin{equation}
  \mn{\EMF}_i = \alpha_{ij} \bar{B}_j 
         - \tilde{\eta}_{ij}\varepsilon_{jkl}\partial_k \bar{B}_l\,,
  \quad i,j \in \left\{R,\phi\right\}, k=z\,,
  \label{eq:param}
\end{equation}
with tensorial coefficients $\alpha_{ij}$ and $\tilde{\eta}_{ij}$. The
determination of these closure parameters \citep{2008AN....329..619G}
hugely benefited from the development of the so-called test field
method \citep{2005AN....326..245S}. For kinematically forced
turbulence, this method allows to determine unambiguously all eight
tensor components -- and, in fact, none of them can be ignored
\citep{2010arXiv1001.5187G}. Notably, we find a significant positive
R{\"a}dler effect in the off-diagonal elements of the diffusivity
tensor (see panel 'd' in Figure~\ref{fig:ae}).

The measured $\alpha$~effect and turbulent diffusion are of the
expected sign and magnitude (see Figure~\ref{fig:ae}, panels 'a' and
'c'). The key finding is that of a moderate downward pumping (panel
'b') as predicted by \citet{1993A&A...269..581R}. This inward pumping
has profound implications in compensating the effect of the equally
strong upward mean flow (not shown), thus yielding ideal conditions
for the dynamo. Moreover, a wind may aid the shedding of small-scale
magnetic helicity as discussed in \citet{2007MNRAS.377..874S}.

\section{Recent advances}

One major concern with the local box approximation was the issue of
its limited dimension in the horizontal direction. In the low-pressure
ambient medium of the galactic halo, supernova remnants can easily
expand to several hundred parsec in diameter. Because of the periodic
boundaries and the finite domain size, this leads to (spuriously)
correlated self-interactions. The simulations of
\citet{1999ApJ...514L..99K} used $L_x=L_y=500\pc$, which was justified
for the limited vertical extent of their model. For our standard runs
we applied $L_x=L_y=800\pc$ together with a vertical box size of $\pm
2\kpc$.

\subsection{Large box simulations}

Even with near $\kpc$ horizontal box size, the issue of
self-interaction was still seen in the far halo region. To eliminate
the possibility of an artificial dependence of the results on the
horizontal box size, we carried out a fiducial simulation run at
$L_x=1.6\kpc$ (cf. Figures~\ref{fig:slices}~\&~\ref{fig:ae}). In the
region of overlap, the extended profiles agree well with their
counterparts derived from the smaller boxes \citep[cf. Fig. 4.5, model
'F4' in][]{2010arXiv1001.5187G}.

Beyond $\pm 2\kpc$, the velocity dispersion increases significantly,
enhancing the turbulent diffusion and resulting in noisy profiles for
the other coefficients. Note, however, the pronounced systematic
$\alpha$~effect in the region of strong vertical gradients near the
midplane (panels 'a' and 'b' in Figure~\ref{fig:ae}). This clearly
supports the paradigm of helicity production being the direct result
of inhomogeneous turbulence \citep{1993A&A...269..581R}.

\subsection{Scaling relations}

Beyond the purpose of mere diagnostics, the profiles shown in
Figure~\ref{fig:ae} serve as a foundation for global large
eddy-simulations solving (\ref{eq:mf_ind}). We believe that the
semi-global approach of vertically stratified local boxes captures the
essential physics behind the turbulent $\alpha$~effect in the galactic
disc. At the same time, we are aware that the local geometry implies
certain restrictions on the permitted dynamo modes. Identifying global
symmetries is however necessary for a direct comparison with
observations.

To warrant global mean-field modelling of the $\alpha\,\Omega$~dynamo,
the scaling of the measured closure coefficients with the relevant
input parameters has to be obtained. This is important because the
angular velocity $\Omega$, the local shear rate $q$, the supernova
rate $\sigma$, and the midplane density $\rho_{\rm c}$ are functions
of the galactocentric distance
\citep{2001RvMP...73.1031F}. \citet{2009IAUS..259...81G} already made
a first step in this direction by inferring the dependence on the
supernova rate. Corresponding mean-field models have been reported in
\citet{2009IAUS..259..467E}. Further scaling relations with respect to
$\Omega$ and $\rho_{\rm c}$ will be reported elsewhere.

First estimates (subject to small number statistics) for a range of
0.1--1 times the galactic star formation rate lead to the relations
\begin{equation}
  C_\alpha \sim \Omega\,H,\quad {\rm and}\quad
  C_\Omega \sim \Omega^{1.5}\,H^2\,\sigma^{-0.5}\,.
\end{equation}
This has the consequence that the dynamo number $D \equiv C_\alpha
C_\Omega \sim \Omega^{2.5} H^3 \sigma^{-0.5}$ scales inversely with
the star formation rate. Therefore we would not expect a stronger
amplification of the large-scale field for stronger star formation
activity. Lastly, the pitch angle $p \simeq \sqrt{C_\alpha/C_\Omega}
\sim \Omega^{-0.25} H^{-0.5} \sigma^{0.25}$, in the kinematic regime,
only seems to depend weakly on the studied parameters. The strongest
dependence here is, however, expected from the local shear rate
$q\equiv{\rm d}\ln \Omega/{\rm d} \ln R$, suggesting a dedicated
parameter survey.


\section{Conclusions and prospects}

Because current numerical simulations are limited to very moderate
Reynolds numbers, it is important to understand how efficient the
observed mechanisms remain under realistic conditions. To achieve
this, it has proven fruitful to study simplified scenarios and run
multiple parameter sets \citep[see][for a comprehensive
review]{2005PhR...417....1B}. Fortunately, the emerging physical
effects are dominated by the outer scale of the turbulence, i.e., as
soon as a rudimentary scale separation is achieved, the turbulent
quantities should become independent of the actual micro scale.

The growing complexity of models challenges the distribution of
computing time: increasing physical realism leaves little margin for
the variation of key parameters, let alone convergence checks or
running multiple representations of a single parameter set. Dedicated
studies remain mandatory to segregate artificial trends from genuinely
physical ones \citep{2009A&A...498..335H,2009IAUS..259...81G}. 

The dilemma becomes even more apparent when looking at the recent
trend to performing ``resolved'' global simulations. For these,
convergence checks are a rare exception. While an ``enhanced''
diffusivity may be sufficiently approximated by the numerical
truncation error on the grid scale, the diamagnetic pumping term
certainly is not. Yet vertical transport has profound implications on
the emerging dynamo modes and growth rates
\citep[see, e.g.,][]{2001A&A...370..635B}.

In conclusion, we advocate a strategy that has been applied with great
success in the design of aircraft, namely the concept of large eddy
(or mean-field) simulations. Global fluid simulations currently cannot
guarantee scale separation for all relevant physical scales.  To
obtain quantitatively correct results, we therefore believe that a
sub-grid scale model is inevitable. Present local box simulations are
a valuable means to provide a rigorous framework for the calibration
of such a model.


\begin{acknowledgments} 
This work used the \textsc{nirvana} code version 3.3 developed by Udo
Ziegler at the Astrophysical Institute Potsdam. Computations were
performed at the AIP \texttt{babel} cluster.
\end{acknowledgments}


\end{document}